\documentclass[12pt]{article}

\def\rdots{\mathinner{\mkern1mu\raise1pt\vbox{\kern1pt\hbox{.}}\mkern2mu
   \raise4pt\hbox{.}\mkern2mu\raise7pt\hbox{.}\mkern1mu}}
\newcommand{\Z}{{\rm Z\kern-.35em Z}}
\newcommand{\bP}{{\rm I\kern-.15em P}}
\newcommand{\Q}{\kern.3em\rule{.07em}{.65em}\kern-.3em{\rm Q}}
\newcommand{\R}{{\rm I\kern-.15em R}}
\newcommand{\h}{{\rm I\kern-.15em H}}
\newcommand{\C}{\kern.3em\rule{.07em}{.65em}\kern-.3em{\rm C}}
\newcommand{\T}{{\rm T\kern-.35em T}}

\newcommand{\be}{\begin{equation}}
\newcommand{\ee}{\end{equation}}

\newcommand{\pa}{\partial}
\newcommand{\cd}{\cdot}

\newcommand{\al}{\alpha}
\newcommand{\nn}{\nonumber}

\begin{document}

 
\openup 1.5\jot
\centerline{A Polymer Expansion for the Quantum}
\centerline{Heisenberg Ferromagnet Wave Function}

\vspace{1in}
\centerline{Paul Federbush}
\centerline{Department of Mathematics}
\centerline{University of Michigan}
\centerline{Ann Arbor, MI 48109-1109}
\centerline{(pfed@umich.edu)}

\vspace{1in}

\centerline{\underline{Abstract}}

A polymer expansion is given for the Quantum Heisenberg Ferromagnet wave
function.  Working on a finite lattice, one is dealing entirely with
algebraic identities; there is no question of convergence.  The
conjecture to be pursued in further work is that effects of large
polymers are small.  This is relevant to the question of the utility of
the expansion and its possible extension to the infinite volume.  In
themselves the constructions of the present paper are neat and elegant
and have surprising simplicity.

\vfill\eject

This paper assumes the fundamentals of the Heisenberg model but is
basically self-contained; it arises from the work in [1], [2], [3], but
these references need not be referred to.  We intend to continue the work
in the present paper, to obtain bounds on polymer contributions enabling
extension to the infinite lattice.  We also have some hope of using this
expansion in a proof of the phase transition.

We work with a finite rectangular lattice, $V$, in $d$-dimensions, $\cal
V$ the set of its vertices.  The Hamiltonian is taken as

\be
H = -\sum_{i \sim j} \: \frac 1 2 \;\Big(\vec \sigma_i \cd \vec \sigma_j
- 1 \Big) = - \sum_{i\sim j} (I_{ij}- 1)
\ee
where $I_{ij}$ interchanges the spins at nearest neighbor sites $i$ and
$j$.  The Hilbert space ${\cal  H}$ is constructed from basis elements
$\vec i_{\cal  S}$, basis elements in $1-1$ correspondence with subsets
${\cal  S}$ of ${\cal  V}$, used for their labeling.  In a spin-up
spin-down representation

\be
\vec i_{{\cal S}} =
\begin{array}[t]{c}
{\displaystyle\otimes} \\
{\scriptstyle {i \in {\cal S}} }
\end{array}
\left( \begin{array}{c}
1 \\
0
\end{array} \right)_i
\begin{array}[t]{c}
{\displaystyle\otimes} \\
{\scriptstyle {j\not\in {\cal S}} }
\end{array}
\left( \begin{array}{c}
0 \\
1
\end{array} \right)_j \;.
\ee
A vector $\vec f$ in ${\cal H}$ may be expanded as
\be
\vec f = \sum_{{\cal S}} f({\cal S}) \vec i_{\cal S} \;.
\ee

For two sets ${\cal S}$ and ${\cal S}'$ we write ${\cal S} \sim {\cal
S}'$ if ${\cal S}'$ is constructed from ${\cal S}$ by replacing some
single element of ${\cal S}$ by one of its nearest neighbors.  That is,
${\cal S} \sim {\cal S} ' $ if there is a set ${\cal F}$ and elements of
${\cal V}$, $i$ and $j$, so that

\begin{eqnarray}
{\cal S} &=& {\cal F} \cup i \nn \\
 \\
{\cal S}' &=& {\cal F} \cup j \nn
\end{eqnarray}
where $i \sim j$ and the unions in (4) are disjoint.  If we write
\be
\vec f(t) = e^{-Ht} \vec f = \sum_{\cal S} f({\cal S}, t) \vec i_{\cal S}
.
\ee
It is easy to see that the $f({\cal S}, t)$ satisfy the differential
equations
\be
\frac \pa {\pa t} \;f({\cal S}, t) = \sum_{{\cal S}' \sim {\cal S}}
\left( f({\cal S}', t) - f({\cal S}, t) \right) \;.
\ee
This is the graph heat equation, corresponding to a graph with vertices
the subsets of ${\cal V}$, and with an edge connecting vertices ${\cal
S}_1$ and ${\cal S}_2$ if and only if ${\cal S}_1 \sim {\cal S}_2$.

We now write ${\cal H}$ as direct sum
\be
{\cal H} = \bigoplus_{n=0}^{\#({\cal V})} \;{\cal H}^n
\ee
where as indicated $n$ ranges from 0 to $\#({\cal V})$.  ${\cal H}^n$ is
spanned by the basis elements $\vec i_{\cal S}$ where $\#({\cal S}) =
n$.  ${\cal H}^n$ is the $n$ spin-wave sector of the Hilbert space ${\cal
H}$.  The ${\cal H}^n$ are invariant subspaces of $H$.  We write $H^n$
for $H$ restricted to ${\cal H}^n$.

We introduce operators $T^{r,s}$, where $T^{r,s}$ is a linear mapping
from ${\cal H}^r$ to ${\cal H}^s$.  They are defined as follows:

\begin{enumerate}
\item[1)] $T^{r,r}$ is the identity on ${\cal H}^r$
\item[2)]  If $s > r$,
\[      T^{r,s} = 0    \]
\item[3)] If $s < r$ let $\vec g$ be in ${\cal H}^r$
\be     \vec g = \sum_{{\cal S}} g({\cal S}) \vec i_{\cal S}    \ee
where $g({\cal S})$ is non-zero only if $\#({\cal S}) = r$.  Let
\be     \vec h = T^{r,s} \vec g = \sum_{\cal S} h({\cal S}) \vec i_{\cal
S}.   \ee
Then $h({\cal S})=0$ unless $\#({\cal S})=s$, and
\be h({\cal S}) = \sum_{{\cal S}' \supset  {\cal S}} g({\cal S}')  \:
{\rm if} \:   \#({\cal S}) = s .  \ee
\end{enumerate}
We note that if $r > s > k$ then
\be   T^{s,k} \; T^{r,s} = \frac{(r-k)!}{(s-k)!(r-s)!} \:T^{r,k}    \ee
This is easy counting.

\bigskip

A nice result is that $T^{r,s}$ intertwines $H^r$ and $H^s$.  That is
\be     T^{r,s} H^r = H^s T^{r,s}   \ee
where both sides of (12) are viewed as mappings form ${\cal H}^r$ to
${\cal H}^s$.  This is treated in Appendix A.  The formalism is from
Section II of [1].  A similar more complex parallel theory is given in
[3] for random walks on the permutation group, instead of subspaces of a
lattice.

We start presenting the polymer expansion for $\vec f(t)$ of equation
(5).  We assume $\vec f(t)$ is normalized so that
\be     \sum_{{\cal S}} f({\cal S}, t) = 1 \:.  \ee
We note that if at any time this equation holds, the heat equation,
equation (6), preserves the identity.  We do not consider the possibility
that the sum on the left side of (13) be zero, so no such normalization
is possible.

We let $\cal P$ be a partition of $\cal V$.  We write ${\cal S}_\alpha <
\cal P$ for a subset ${\cal S}_\al$ of the partition $\cal P$.  One has

\begin{eqnarray}
S_\al \bigcap S_\beta &=& \emptyset,   \al \not= \beta \\
\bigcup_{\al \in I^P} {\cal S}_\al &=& {\cal V} .
\end{eqnarray}
We will have
\be \vec f(t) = \sum_{{\cal P}}
\begin{array}[t]{c}
{\displaystyle\otimes} \\
{\scriptstyle {\cal S}_\al <{\cal P}}
\end{array}
\vec u({\cal S}_\al, t)
\ee
where
\be
\vec u({\cal S}_\al, t)= \left(
\begin{array}{c}
\phi_i(t) \\
1-\phi_i(t)
\end{array}
\right)_i
\ee
if ${\cal S}_\al = \{ i\}$.

If $\#({\cal S}_\al) = r > 1$
\be
\vec u({\cal S}_\al, t) = u^r ({\cal S}_\al, t)
\begin{array}[t]{c}
{\displaystyle\otimes} \\
{\scriptstyle {i \in {\cal S}_\al } }
\end{array}
\left( \begin{array}{c}
1 \\
-1
\end{array} \right)_i
\ee
We also write
\be
u({\cal S},t) = \left\{
\begin{array}{lll}
\phi_i(t) \:\:&{\rm if}& \:\:{\cal S} = \{i\} \\
u^r({\cal S},t) &{\rm if}& \:\:\#({\cal S}) = r > 1
\end{array} \right. .
\ee
We write $\vec f(t)$ as a sum of its different spin-wave number
components
\be     \vec f(t) = \sum^{\#({\cal V})}_{n=0} \vec f_n(t)       \ee
\be     \vec f_n(t) \in {\cal H}^n      \ee
We set
\be     \vec c_r(t) = \sum^{\#({\cal V})}_{n=0} T^{n,r} \vec f_n(t)    
\ee
and
\be     \vec c_r(t) = \sum_{{\cal S}} c^r({\cal S},t) \vec i_{\cal S}  
\:.  \ee
(Do notice that the $c^{r},s$ satisfy the graph heat equation, (6).)
Then we find that equation (16) is satisfied if the  $u({\cal S},t)$ are
chosen to satisfy:
\be
c^r({\cal S},t) = u^r({\cal S}, t) + \sum_{{\cal P}}
\begin{array}[t]{c}
{\displaystyle\bigotimes} \\
{\scriptstyle {\cal S}_\beta <{\cal P}}
\end{array}
u ({\cal S}_\beta, t)
\ee
Where here $\cal P$ is a ${\bf proper}$ partition of $\cal S$ and
$\#({\cal S}) = r$.  $r$ will range from 1 to $\#(\cal V)$.  Equations
(16) and (24) are prototype cluster-expansion/polymer-expansion
equations.  But the form of equation (18) is perhaps surprising? 
Appendix B treats the consistency of the formalism; that there is a
unique solution for the $u's$ from (24), and they yield equation (16).

\vfill\eject
\noindent
\underline{Appendix A.  Intertwining Result}

In virtue of equation (11) it is enough to show $T^{r,r-1}$ intertwines. 
We choose to show equivalently that $T^{r,r-1}$ carries a solution of the
heat equation into a solution of the heat equation.   Let $f({\cal S},t)$
satisfy the heat equation, and be zero unless $\#({\cal S}) = r$.  We
define

$$      g({\mathit s} , t) = \sum_j f({\mathit s} \cup j, t),
\:\#({\mathit s}) = r-1  \eqno(A.1) $$
We wish to show $g$ satisfies the heat equation.  Writing the heat
equation for $f$:
$$\frac{\pa f}{\pa t} (s \cup i, t) = \sum_{{\cal S}' \sim(s \cup i)}
\left(f ({\cal S}', t) - f(s \cup i,t)\right) \eqno(A.2)
$$
We sum the two sides of (A.2) over $i$.
$$\frac{\pa}{\pa t}g (s, t) = \sum_i \sum_{{\cal S}' \sim(s \cup i)}
\left(f ({\cal S}', t) - f(s \cup i,t)\right) \eqno(A.3)
$$
The right side splits into two terms $I_1$ and $I_2$
$$   I_1 = \sum_i \sum_{ s' \sim  s} \left(f ( s' \cup i, t) - f(s \cup
i,t)\right) \eqno(A.4)  $$
and
$$   I_2 = \sum_i \sum_{j \sim i} \left(f ( s \cup j, t) - f(s \cup i,
t)\right) \eqno(A.5)  $$
It is easy to see
$$   I_1 =  \sum_{s' \sim s} \left(g (s' , t) - g(s ,t)\right)
\eqno(A.6)  $$
and just a little harder to see
$$    I_2 = 0   $$
and the result is proved.

\noindent
\underline{Appendix B.  In Partes Tres}.

We divide the demonstration of consistency into three parts.

I)  We first note that equation (24) has a unique solution for the $u^r$
(these are the unknowns).  One solves inductively over $r$, the $r$th
equation uniquely determining $u^r$.

II)  Once the $u$'s are determined from equation (24), we substitute them
in the right side of equation (16) which we call $\vec X(t)$, so equation
(16) becomes
$$   \vec f(t) = \vec X(t).     \eqno(B.1)      $$
(Of course we do not know whether (B.1) is true, that is what we're
trying to show.)  We decompose $\vec X(t)$
$$      \vec X(t) = \sum^{\#({\cal V})}_{n=0} \vec X_n(t)   \eqno(B.2)  
$$
$$      \vec X_n(t) \in {\cal H}^n   \eqno(B.3)   $$
and define
$$       \vec d_r(t) = \sum^{\#({\cal V})}_{n=0} T^{n,r} \vec X_n(t)  
\eqno(B.4)   $$
$$       \vec d_r(t) = \sum_{\cal S} d^r({\cal S}, t) \vec i_{\cal S}
\:.  \eqno(B.5)   $$
The result we seek to now show is the following:  If $d^r({\cal S}, t) =
c^r({\cal S}, t)$ all ${\cal S}, r$ then $\vec f(t) = \vec X(t)$.

This we also show by induction over $r$, but in the opposite direction,
from $r = \#({\cal V})$ down to $r=0$.  At the step $r=r$ we clearly get
$$      \vec f_r(t) = \vec X_r(t) \:.  \eqno(B.6)       $$
(One only needs $T^{r,r} = I$, and $T^{r,s} = 0$ if $s > r$.)

III)    We are left with the task of showing
$$      d^r({\cal S},t) = c^r({\cal S},t) \:. \eqno(B.7)        $$
We first do a preliminary investigation.

Let
$$      \vec h(t) = \sum h({\cal S}, t) \vec i_{\cal S}    \eqno(B.8)  $$
$$  = \sum^{\#({\cal V})}_{n=0} \vec h_n(t)     \eqno(B.9)      $$
$$  \vec h_n(t) \in {\cal H}^n       \eqno(B.10)   $$
and define
$$    \vec g_r(t) = \sum_n T^{n,r}\; \vec h_n(t) \eqno(B.11) $$
$$  =  \sum_{{\cal S}} g^r({\cal S},t) \vec i_{\cal S} \eqno(B.12)  $$
We then find the following expression for $g^r({\cal S}, t)$
$$ g^r({\cal S}, t) = \begin{array}[t]{c}
{\displaystyle\sum} \\
{\scriptstyle{\cal S}'} \\
{\scriptstyle {\cal S}' \cap {\cal S} = \phi}
\end{array}
h({\cal S} \cup {\cal S}', t)  \eqno (B.13)
$$
where $\#({\cal S}) = r$.

Now when we compute $d^r({\cal S},t)$ using the expression (B.13) with
$X$ replacing $h$ $(X(t) = \sum_{\cal S} X({\cal S},t) \vec i_{\cal S})$,
the only terms in the expression for $\vec X(t)$ from (16) which will
contribute are of the form

$$ \left( \vec u^r({\cal S},t) + \sum_{\cal P} \bigotimes_{{\cal S}_\beta
< \; {\cal P}} \vec u({\cal S}_\beta,t) \right)
\bigotimes_{i \not\in {\cal S}}
\left( \begin{array}{c}
\phi_i(t) \\
1 - \phi_i(t)
\end{array} \right)_i
\eqno(B.14)    $$
using the notation from equation (24).  That is because the sum over
${\cal S}'$ in (B.13) may be written as an iterated sum, summing for each
vertex not in ${\cal S}$, whether the vertex is in ${\cal S}'$ or not. 
This amounts to summing over spin-up and spin-down at that vertex.  At
vertex $k$ this sum applied to the term in the tensor product

\[    \left( \begin{array}{c}
\phi_k(t) \\
1 - \phi_k(t)
\end{array} \right)_k  \]
yields 1, and applied to
\[      \left( \begin{array}{r} 1 \\ -1 \end{array} \right)_k   \]
yields 0.  We get from the terms in $\vec X(t)$ in (B.14) that
$$   d^r( {\cal S},t) = c^r({\cal S},t) \eqno(B.15)   $$
Quod erat demonstrandum.

\bigskip
\bigskip
\noindent
\underline{Acknowledgment}:  I would like to thank the referees from the
{\it Letters in Mathematical Physics} for encouraging me to write this
paper in a more intelligible form.

\bigskip
\bigskip

\centerline{\underline{References}}

\begin{itemize}
\item[[1]] P. Federbush, ``For the Quantum Heisenberg Ferromagnet, Some
Conjectured Approximations", math-ph/0101017.
\item[[2]]  P. Federbush, ``For the Quantum Heisenberg Ferromagnet, A
Polymer Expansion and its High T Convergence", math-ph/0108002.

\item[[3]] Robert T. Powers, ``Heisenberg Model and a Random Walk on the
Permutation Group", {\it Lett. in Math. Phys.} {\bf 1}, 125-130 (1976).

\end{itemize}

\end{document}